\documentclass[12pt]{iopart}
\usepackage{epsfig}  
\begin{document}

\title[Correlations and Fluctuations]{Correlations and Fluctuations}

\author{Harald Appelsh\"{a}user }

\address{Gesellschaft f\"{u}r Schwerionenforschung, Planckstrasse 1,
64220 Darmstadt, Germany}

\begin{abstract}
New results on particle correlations and event-by-event fluctuations
presented at Quark Matter 2004 are reviewed. 
\end{abstract}



%\maketitle

%\section{Introduction}
%Most of our present knowledge about the bulk properties of
%ultra-relativistic heavy-ion
%collisions is based on the study of single-particle inclusive
%distributions. 
%Studies of particle correlations
%and event-by-event fluctuations provide additional information which 
%cannot be obtained from single-particle inclusive observables.
%At this conference, a lot of new correlation and
%fluctuation data have been shown which add to
%the existing beam energy
%and system size systematics, but also a number of
%new observables have been presented.
%The current status is briefly reviewed below.
 
%\section{Fluctuations of the mean transverse momentum}
\section{Event-by-event fluctuations}
Non-statistical event-by-event fluctuations
%of the mean transverse momentum $M_{p_t}$ 
have been 
proposed as a possible signature for the QCD phase 
transition~\cite{stephanov,asakawa}.
The passage of the system close to the critical
point of the QCD phase diagram might be indicated by
a non-monotonic evolution of 
%$M_{p_t}$ 
fluctuations
as function of beam energy. These exciting predictions
triggered an extensive study of 
%$M_{p_t}$ 
event-by-event fluctuations at SPS and at RHIC.

\subsection{Fluctuations of the mean transverse momentum}
\begin{figure}
\begin{center}
%\epsfbox{file.eps}
\mbox{\epsfig{file=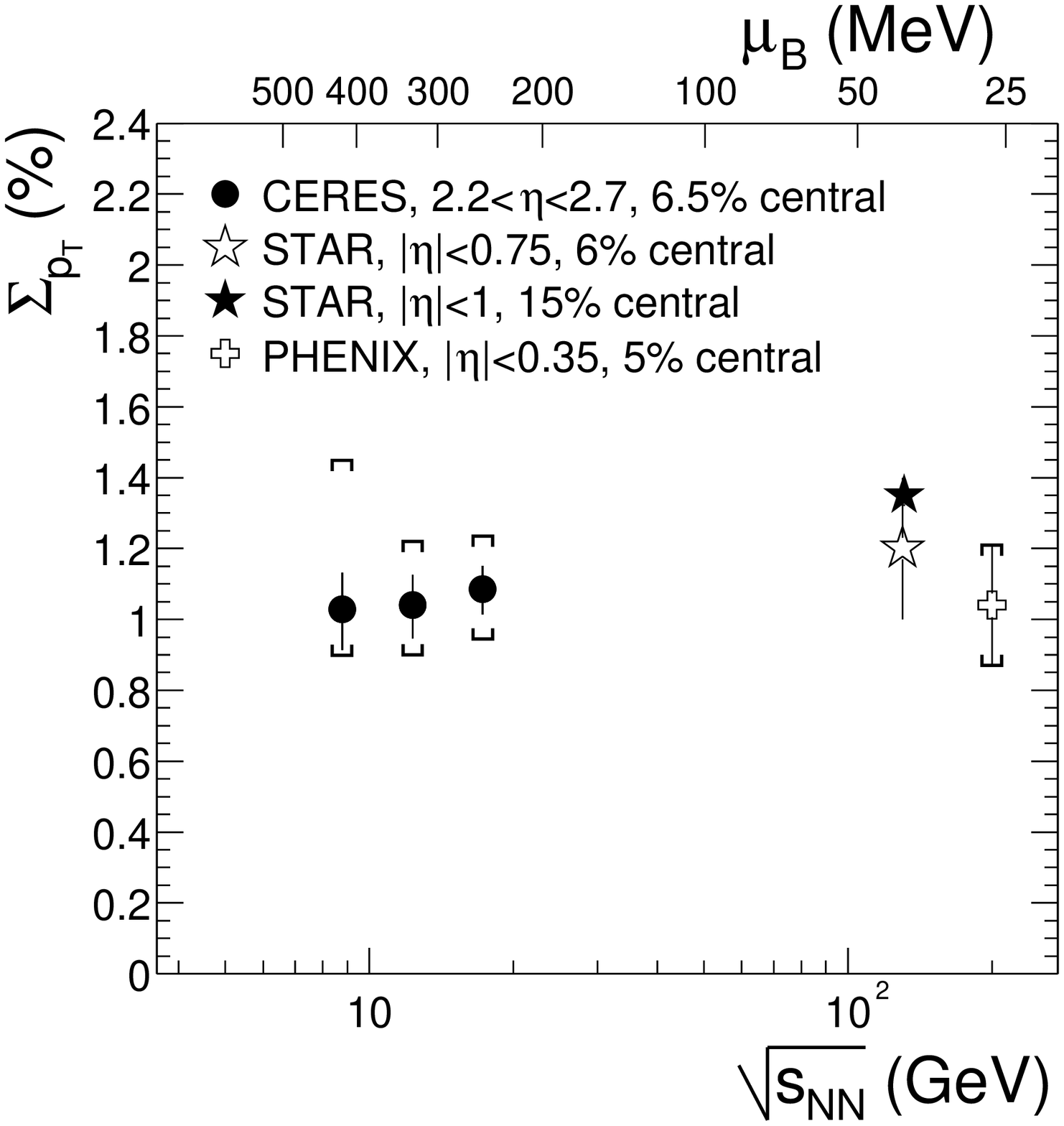,height=7.5cm}}
\mbox{\epsfig{file=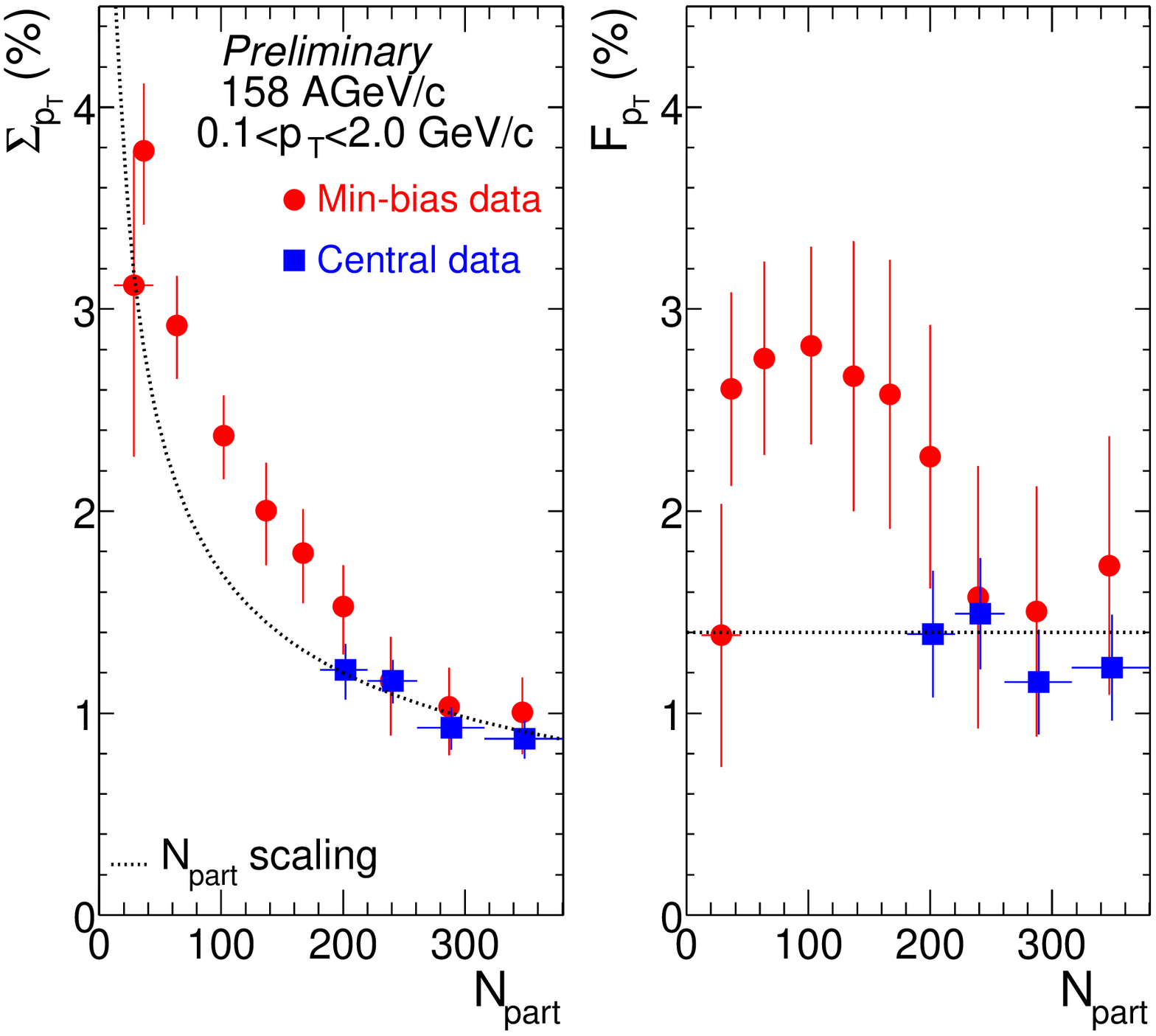,height=7.5cm}}
\caption{Left panel: Beam energy dependence of $M_{p_t}$ fluctuations
in central Pb-Au and Au-Au collisions, 
presented by the CERES collaboration~\cite{hiro}. 
The quantity $\Sigma_{p_t}$ measures the strength
of non-statistical fluctuations in percent of $\langle p_t \rangle$.
Middle and right panel: centrality dependence of $\Sigma_{p_t}$
and $F_{p_t}$ in Pb-Au at 158~AGeV~\cite{hiro}. 
The extrapolation from p-p is indicated by the
dotted line.}
\label{fig1}
\end{center}
\end{figure}

The existence of non-statistical event-by-event fluctuations
of the mean transverse momentum $M_{p_t}$ at 
SPS and RHIC is by now well established. However,
these dynamical fluctuations are small in central collisions, 
typically about 1\% of 
the inclusive mean transverse momentum $\langle p_t \rangle$,
and  only weakly depending on $\sqrt{s_{NN}}$,
see Fig.\ref{fig1} (left panel)~\cite{hiro, ceres_ebe}. 
This value is roughly compatible with the extrapolation
of fluctuation measurements in p-p collisions~\cite{braune} 
under the assumption of 
an independent superposition of particle sources in A-A. 
In particular, no indication for a 
non-monotonic beam energy dependence has been found so far.

Despite the absence of a 'smoking gun' signature for the phase
transition or the critical point, the systematic study
of $M_{p_t}$ fluctuations gives valuable insight into the
particle production mechanism and the dynamic evolution of
the system which cannot be extracted from inclusive distributions.
A quantitative study requires 
an appropriate formalism which facilitates a comparison of
results among different experiments and to theory. 
Presently, an unfortunate situation
has arisen: Practically each experiment uses different
measures for fluctuations, see~\cite{ceres_ebe,phidef,phenix_pap,star_lett,
volo,mitchell}. 
These measures have very
different sensitivities to particular experimental conditions,
such as track quality cuts, tracking efficiency, and acceptance.
In this sense, measures which are most closely related to single
and two-particle densities appear preferable since they are the
least sensitive to trivial efficiency effects~\cite{volo}. 
In this situation, it is mandatory that experiments provide 
all the information necessary for
an approximative conversion of one measure into another.

It was also pointed out that there can be non-trivial
acceptance effects~\cite{westfall}. 
Depending on the scale of the underlying
correlation, the measured fluctuation pattern may change 
as function of the acceptance window. 
For instance, back-to-back configurations may enhance flow or 
jet-like correlations. An attempt for a 
'differential', scale-dependent
analysis of $M_{p_t}$ fluctuations has been presented at this
conference~\cite{trainor}.

Attention has been raised recently by the centrality dependence of
$M_{p_t}$ fluctuations observed at SPS and RHIC. Experiments
NA49, CERES, PHENIX, and STAR observe $M_{p_t}$ fluctuations
which are significantly increased over the p-p extrapolation
in semi-central events~\cite{hiro,star_lett,westfall,kasia,na49_cent,
tannenbaum,phenix_lett}
(e.g. middle and right panel in Fig.\ref{fig1}). 
PHENIX attributed the non-monotonic
centrality dependence of the measure $F_{p_t}$~\cite{phenix_lett} 
to jet production
in peripheral events, combined with jet suppression in more
central events, causing a decrease of 
fluctuations~\cite{mitchell,tannenbaum,phenix_lett}. It was,
however, argued that the redistribution of the quenched jet
energy to lower transverse momenta may not be treated 
consistently in their model. Gavin interpreted the centrality
dependence in terms of thermalization~\cite{gavin,gavin_prl}, 
but a reasonable description
of the data was also found in the framework of a 
string percolation model~\cite{ferreiro,ferr}. 
Although quite different in the 
underlying physical picture, the thermalization and the
percolation approach have an important feature in common:
The well-established increase of 
mean $p_t$ as function of centrality does not reflect 
a smooth excitation of transverse phase space, but rather 
results from a superposition of distinct $p_t$ scales.  
A lower one, characteristic for elementary p-p collisions,
and a larger one established in thermalized 'clumps'
or string clusters. As a consequence, largest fluctuations 
occur in semi-central events where spatial inhomogeneities
are maximal and both scales contribute with about equal strength.

\subsection{Fluctuations of net charge}
%\begin{figure}
%\begin{center}
%\mbox{\epsfig{file=eta_nudyn.eps,height=5.9cm}}
%\hspace{-0.5cm}
%\mbox{\epsfig{file=Npart_nudyn.eps,height=5.9cm}}
%\caption{Left panel: The net fluctuation measure $\nu_{+-,dyn}$,
%normalized to the result at $|\eta|=1$,
%as function of the pseudorapidity bin size. Central Au-Au events
%at different beam energies are compared to p-p. Right panel:
%Centrality dependence of $N_{\rm part}\cdot\nu_{+-,dyn}$ for 
%different beam energies. The data are from STAR~\cite{westfall}.}
%\label{fig2}
%\end{center}
%\end{figure}

%{\em Fluctuations of net charge.}
Results on net charge fluctuations at SPS and RHIC have
been presented at this conference~\cite{hiro,mitchell,westfall,pruneau,christ}. 
All experiments 
report that net charge fluctuations are smaller than
the statistical expectation for independent particle
emission. To a large extent, these deviations can be
attributed to global charge conservation. Additional 
contributions are small at lower SPS energies~\cite{hiro,christ} but possibly
increase with beam energy~\cite{hiro,westfall,star_cfluc}. 
These could be subject to local
charge correlations, e.g.~due to resonance decays. Indeed,
resonance gas models are able to reproduce the measured
net charge fluctuations reasonably well~\cite{hiro}. 
Particle production via resonances (or more general: clusters) 
exhibits a characteristic correlation scale of about one unit
in rapidity~\cite{bell}. STAR has measured the net charge fluctuation
$\nu_{+-,dyn}$ as function of the integrated pseudorapidity 
range~\cite{westfall,pruneau,star_cfluc}.
They reported that $|\nu_{+-,dyn}|$ decreases monotonically
as the rapidity window is opened, in line with the above mentioned
correlation scale due to particle production in clusters.
This pattern may, however, be modified by dynamics
and the details of the space-time evolution. Indeed, STAR
observes that the (relative) $\eta$ dependence of $\nu_{+-,dyn}$
in peripheral Au-Au collisions at $\sqrt{s}=20$, 130, and 200~AGeV
is very similar to p-p, while in central Au-Au collisions
the fluctuation signal is significantly more focussed at small 
pseudorapidity
windows.
%, see left panel in Fig.~\ref{fig2}. 
As pointed out in~\cite{volo,pruneau},
this effect is essentially equivalent with the narrowing
of the balance function~\cite{pratt}, observed by STAR and 
NA49~\cite{westfall,christ,marek,bfun_star},
and may be related to an increase of $\langle p_t \rangle$
in central collisions. Another important implication of these
results is that hadronization occurs "late", i.e. that diffusive
processes after hadron production may not be very effective,
in line with the notion of a short-lived hadronic phase.

%Detailed studies of the centrality dependence of net charge
%fluctuations have been presented~\cite{hiro,westfall,star_cfluc}. 
%A small but significant 
%deviation from a simple $\langle N \rangle$ or $N_{\rm part} $
%scaling behavior has been observed in semi-peripheral 
%events (Fig.\ref{fig2}, right panel). 
%This effect could be attributed
%to thermalization, as proposed by Gavin~\cite{gavin}, but
%future studies and a more detailed understanding of 
%correlated particle production in p-p and A-A are needed.

PHOBOS has measured charge-independent forward-backward
multiplicity fluctuations~\cite{wozniak}. Fluctuations in 
their observable 
$N_{\rm forw}-N_{\rm backw}$ are enhanced
with respect to the statistical expectation,
depending on the size and the separation of the forward
and backward pseudorapidity intervals. These results 
point to correlated particle production which extends
over about one unit of rapidity, qualitatively consistent
with the net charge fluctuation results described before.

\subsection{Multiplicity and particle ratio fluctuations}
\begin{figure}[t]
\begin{center}
\mbox{\epsfig{file=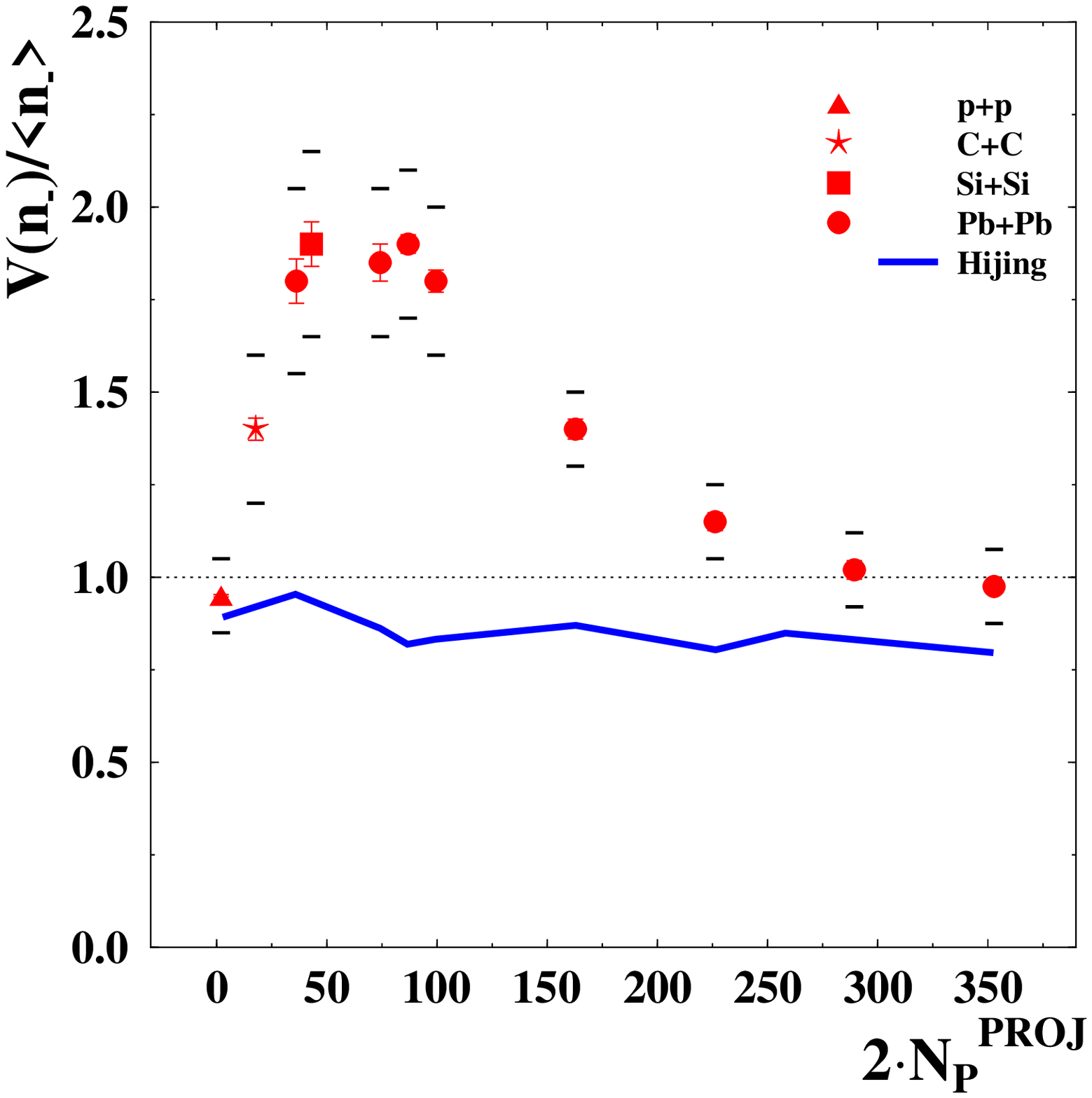,height=7cm}}
\mbox{\epsfig{file=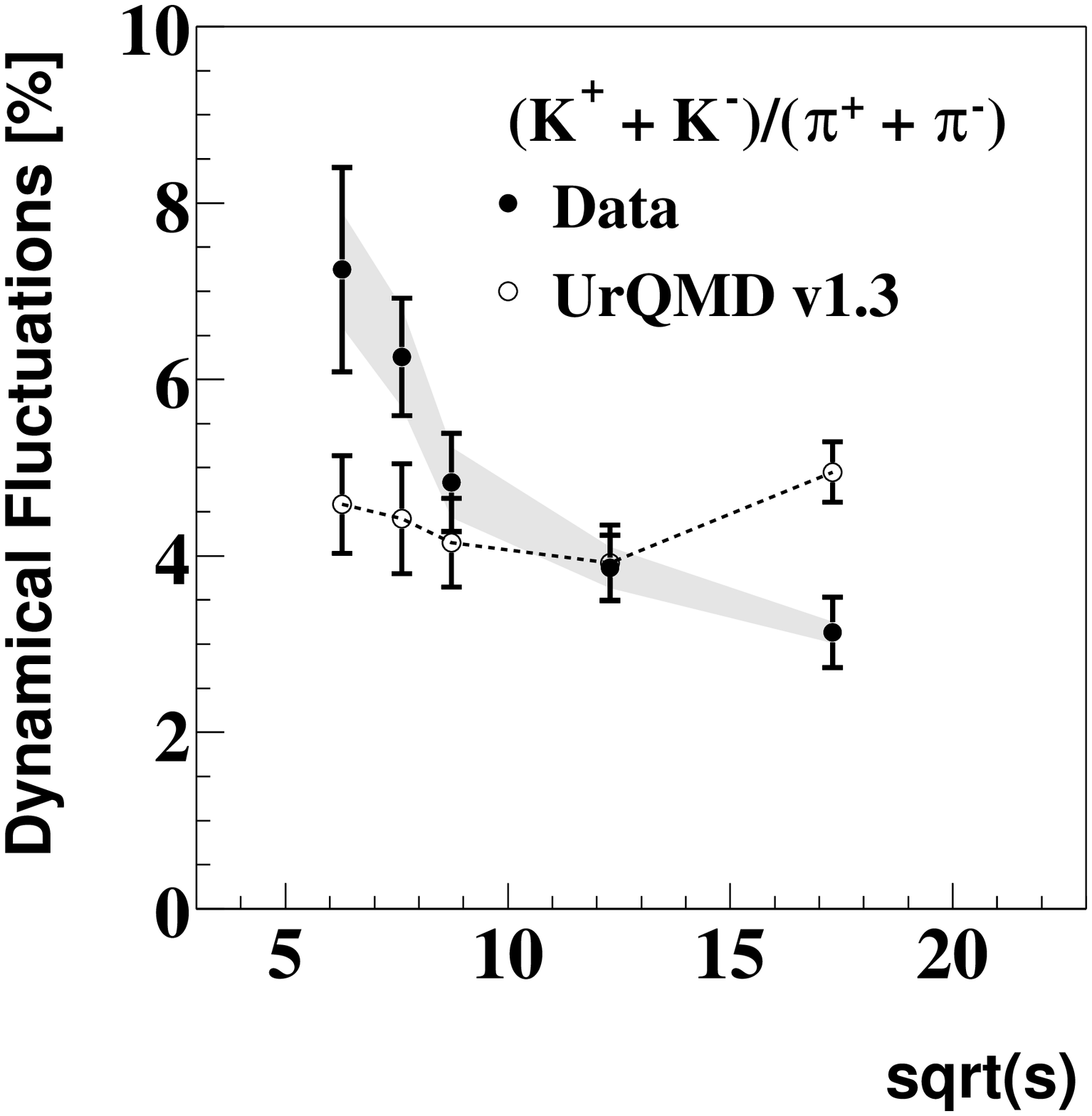,height=7.2cm}}  
\end{center}
\caption{Left panel: Fluctuations of negatively charged particle
multiplicity at 158 AGeV as function of the number of projectile
participants, presented by NA49~\cite{marek}. 
Right panel: NA49 results on fluctuations of the ratio 
$(K^++K^-)/(\pi^++\pi^-)$ in central Pb-Pb as function of
energy~\cite{roland}.}
\label{figna49}
\end{figure}

%{\em Multiplicity and particle ratio fluctuations.}
Experiment NA49 has, for the first time, presented results
on event-by-event fluctuations of the negatively
charged particle multiplicity at 158 AGeV/c~\cite{marek}. 
As a function of the number of projectile participants,
a non-monotonic behaviour with a distinct maximum in
intermediate mass systems is observed, see Fig.~\ref{figna49} (left panel). 
Note the similarity
to the centrality dependence of $M_{p_t}$ fluctuations (Fig.~\ref{fig1}).
By the definition of $M_{p_t}$, a connection between multiplicity and 
$M_{p_t}$ fluctuations
is not unexpected. Future studies could show 
to what extent these two quantities fluctuate in a correlated
way.

Event-by-event fluctuations of the particle yield
ratios $(K^++K^-)/(\pi^++\pi^-)$
and $(p+\bar{p})/(\pi^++\pi^-)$ have been studied by NA49 as 
a function of beam energy~\cite{roland}. Fluctuations of the 
$(p+\bar{p})/(\pi^++\pi^-)$-ratio 
are consistent with UrQMD calculations 
and most likely due to correlations from
resonance decays. In contrast, $(K^++K^-)/(\pi^++\pi^-)$ fluctuations 
show a significant energy dependence and are enhanced with respect to
model calculations, as demonstrated in the right panel of 
Fig.~\ref{figna49}. The enhancement is most pronounced at the lowest
beam energy. These preliminary
results might be connected with the previously reported 
non-monotonic behaviour of the $K^+/\pi^+$ excitation function~\cite{marek} 
and
could give insight with regards
to the possible occurence of the phase transition.
Further investigations, in particular of charge-dependent fluctuations,
but also more extended model studies
are needed to unravel trivial and non-trivial 
contributions to this observable. 

\section{Two-particle correlations}

Previous HBT studies of identical pions in Au-Au collisions 
at RHIC yielded source parameters only 
moderately larger than at lower beam energies.
In particular, hydrodynamic models, successful at 
RHIC in describing single particle momentum space
distributions, fail to reproduce the measured HBT
radii (see~\cite{magestro} for an overview). 
The observed evolution and emission
time scales are apparently shorter than predicted 
by the models.
This so-called HBT-puzzle has not yet been entirely
resolved. In fact, more puzzling HBT results have
been presented at this conference.

\subsection{Pion-HBT: Beam energy and system size dependence}

\begin{figure}[t]
\begin{center}
\mbox{\epsfig{file=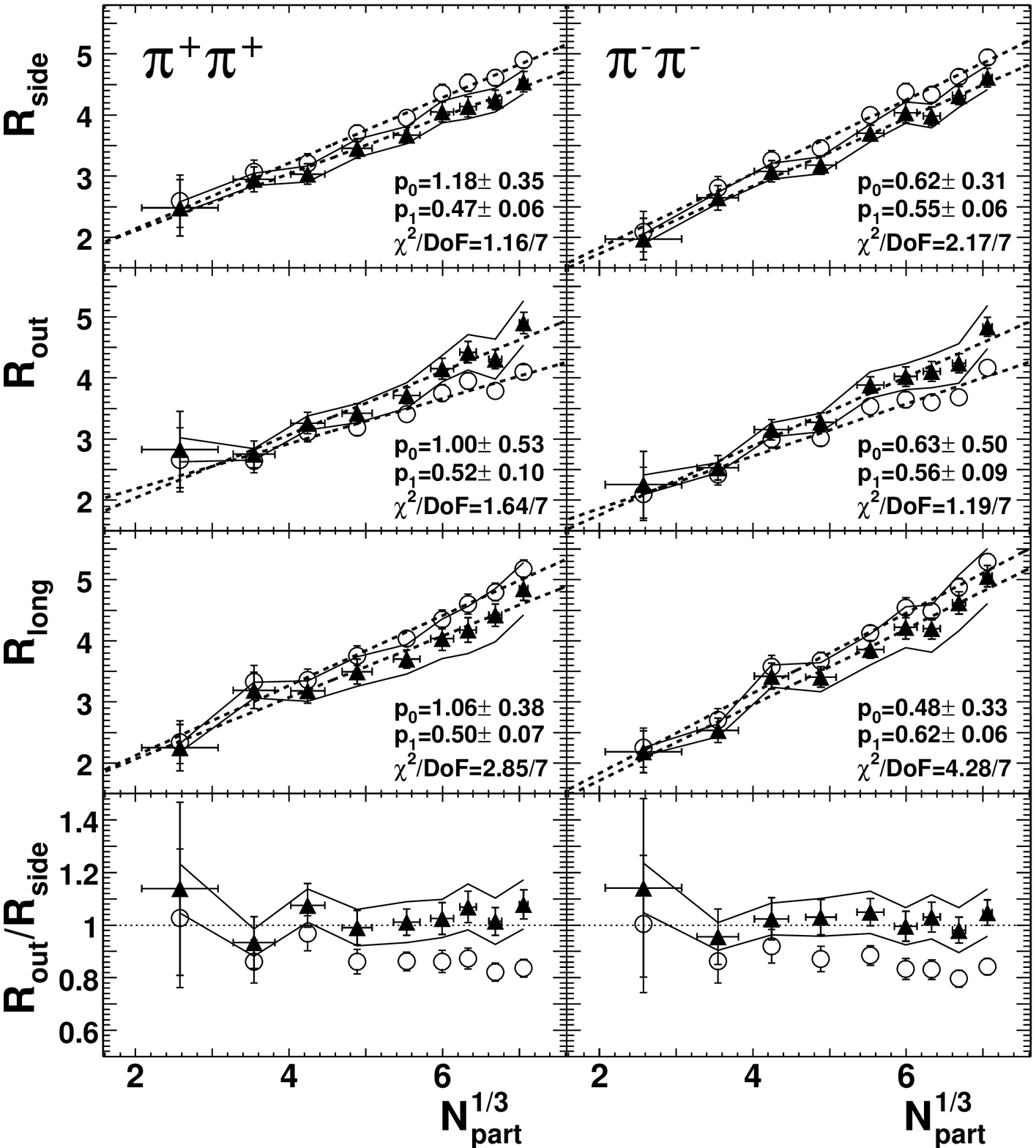,height=8.5cm}}
\hspace{-0.0cm}
\mbox{\epsfig{file=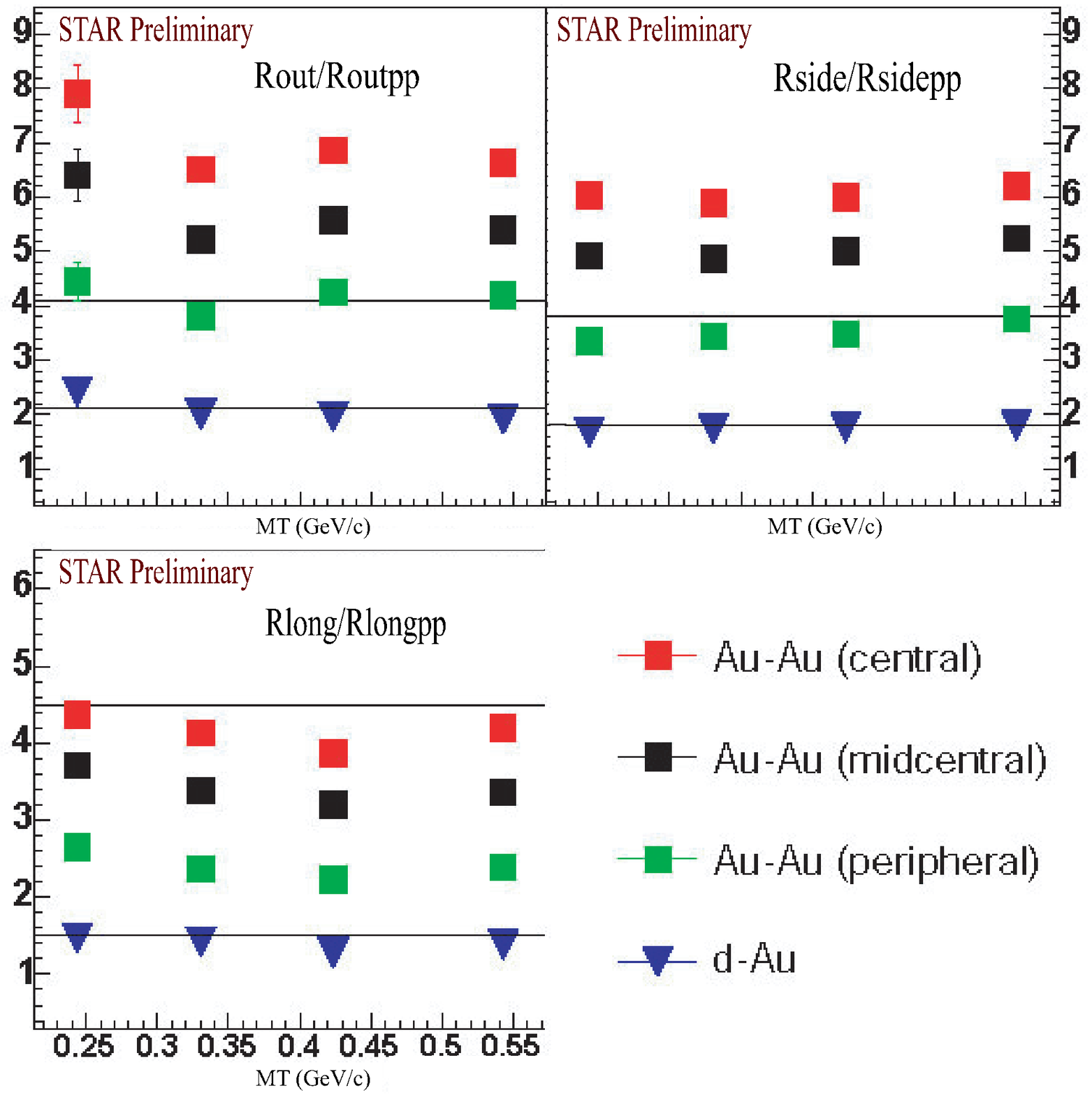,height=7.5cm}}  
\caption{Left panel: Centrality dependence of identical-pion HBT 
results from Au-Au at 200 GeV by PHENIX. Results employing the
new Coulomb treatment (full symbols) are compared to the
traditional approach
assuming a pure pion sample (open symbols)~\cite{heffner,phenix_newhbt}. 
Right panel: Ratio of HBT radii in d-Au and Au-Au at 200 GeV
with respect to p-p as function of $k_t$, presented by 
STAR~\cite{magestro,gut,bekele}.}
\label{fig_HBT}
\end{center}
\end{figure}

I am happy too see
that by now all experiments adopted a new treatment of
the final state Coulomb repulsion which takes into account the 
finite purity of the pion sample~\cite{sinyu}. 
In Fig.~\ref{fig_HBT} (left panel)~\cite{heffner}
the results of the new treatment (full symbols) are compared
to those assuming a pure pion sample (open symbols). As 
demonstrated earlier~\cite{cereshbt} the HBT radii,
in particular
the ratio $R_{\rm out}/R_{\rm side}$, 
are very sensitive in this respect. 

PHENIX presented a very detailed centrality dependence
of pion HBT radii in Au-Au at 200 GeV~\cite{heffner,phenix_newhbt}
(Fig.~\ref{fig_HBT}, left panel). 
All three Bertsch-Pratt
parameters show a linear increase with $N_{\rm part}^{~~1/3}$.
In some ways, this is a surprising result: the space-time properties
and the freeze-out
dynamics are expected to change drastically from very peripheral
to central events. The HBT radii, however, 
show no indication of a qualitative change. They scale 
essentially with the initial volume of the participating 
nucleons. 

A detailed energy scan of identical-pion
HBT at SPS was presented by NA49~\cite{kniege}. 
These preliminary results show only a 
very weak beam energy dependence. 
Inconsistencies with existing HBT data from SPS~\cite{cereshbt},
most pronounced in $R_{\rm out}$, require further
investigations.

The observed beam energy dependence of HBT radii had 
been shown to be consistent with a universal mean free
path at freeze-out $\lambda_f\approx 1$~fm~\cite{ceres_uni}. 
Magestro showed that
this universality holds also in p-p and d-Au at RHIC~\cite{magestro}.
The idea of a constant $\lambda_f$ independent
of the system size seems to disagree with the notion
of a rescattering-dominated late stage of the collision.
Although the system size increases, freeze-out remains
a local phenomenon. 
It is conceivable that an expected increase of $\lambda_f$
with system size is counterbalanced by an increase of
the expansion rate, however, a quantitative cancellation over the
full centrality range is astonishing.

%A possible explanation could be an accidental cancellation 
%due to increasing (transverse) collectivity. 

The dependence of the HBT radii on the mean pair 
transverse momentum $k_t$ is expected to be sensitive
to the details of the dynamical evolution.
New results on $k_t$-dependences have been shown
by NA49~\cite{kniege}, PHENIX~\cite{heffner,phenix_newhbt},
PHOBOS~\cite{phobos_hbt}, and STAR~\cite{magestro,gut,bekele}.
STAR presented a compilation of pion HBT radii from p-p,
d-Au, and Au-Au at 200 GeV.
The $k_t$-dependences of the HBT radii indicate strong
space-momentum correlations in all systems.
In central Au-Au collisions, such space-momentum correlations
had been attributed to hydrodynamic flow. In elementary
collisions, space-momentum correlations arise from
string fragmentation. Also resonance decays may play 
an important role in p-p. It is, however, striking that the
shape of the $k_t$-dependences does not change significantly with 
system size, as shown in the right panel of Fig.~\ref{fig_HBT}, 
although the underlying mechanisms are 
presumably very different.  
This result clearly disfavors an 
independent superposition of p-p: The pronounced 
$k_t$-dependences observed in Au-Au indicate space-momentum 
correlations over a range which is of the order of the system size. 
In the
case of independent p-p collisions the range of space-momentum
correlations would be confined to the size of individual
nucleons, and result in significantly weaker 
$k_t$-dependences~\cite{lisapriv}.
In this sense, the HBT data reveal characteristic features of 
self-similarity between p-p and Au-Au, rather than a simple
superposition.

In essence, system size dependent HBT studies at RHIC
reveal new aspects of the HBT-puzzle: At fixed beam
energy, HBT radii scale with the initial reaction volume.
The evolution of the $k_t$-dependences with system size
exhibit a self-similar pattern and do not hold with an
independent superposition of p-p collisions. 
However, no qualitative change of the dynamical
evolution from p-p to central Au-Au is observed. 
Possible conclusions are surprising: Either
the space-time evolution in central Au-Au is very
similar to that in p-p, or the sensitivity of interferometric
measurements to the dynamics of the system is not as 
expected~\cite{wong,kapusta}. 
The answer to this problem
remains one of the big challenges of this field. 

\subsection{Azimuthally-sensitive HBT studies}

A study of HBT-parameters relative to the reaction
plane in non-central collisions allows the reconstruction
of the spatial source anisotropy at freeze-out~\cite{heinz}.

\begin{figure}[t]
\begin{center}
\begin{minipage}[t]{7 cm}
\epsfig{file=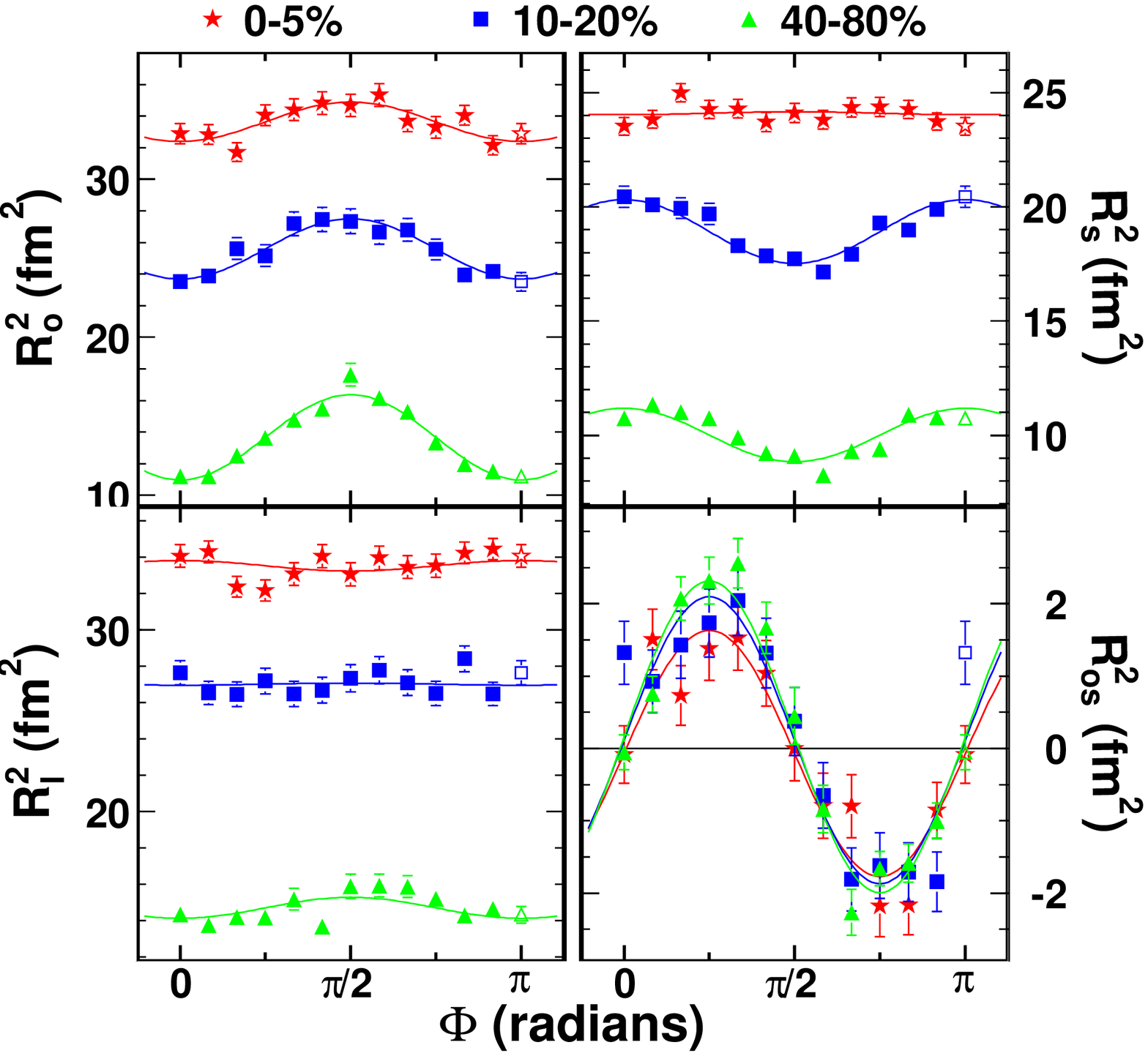,scale=0.45}
\end{minipage}
\begin{minipage}{8 cm}
\vspace{-7.8cm}
\hspace{2cm}
\epsfig{file=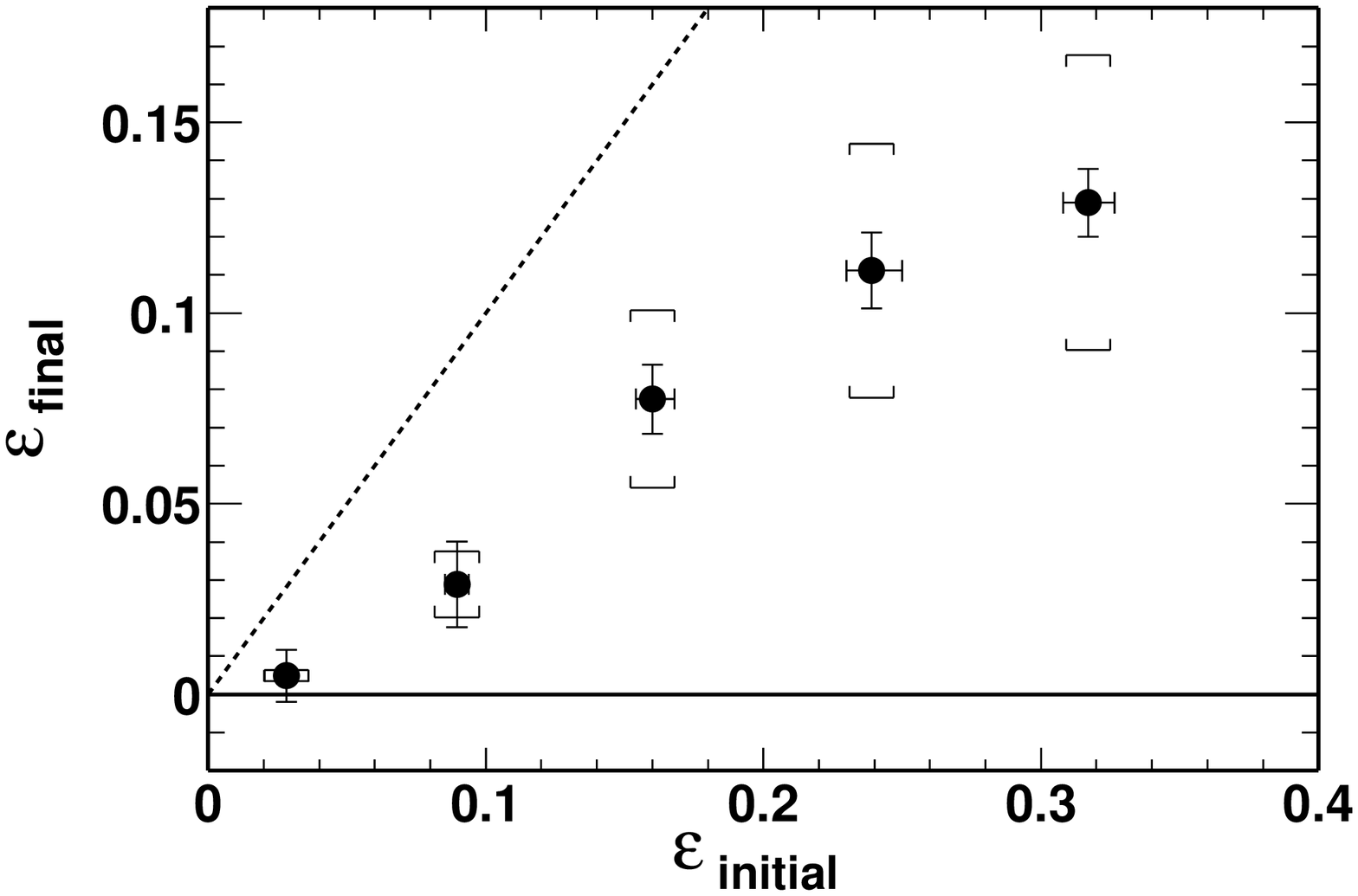,scale=0.3}

\vspace{-0.5cm}
\hspace{-7cm}
\caption{Left panel: HBT radii from Au-Au at 200 GeV 
with respect to the reaction plane and for different 
centralities, presented by STAR. 
Right panel: Reconstructed source eccentricity $\epsilon_{\rm final}$
at freeze-out as function of the initial eccentricity
$\epsilon_{\rm initial}$~\cite{magestro,bekele,azipap}.}
\label{fig_AZI}
\end{minipage}
\end{center}
\end{figure}

Results from a new analysis by STAR in non-central Au-Au collisions
at 200 GeV~\cite{magestro,bekele,azipap} (Fig.~\ref{fig_AZI}, left panel)
confirm earlier
conclusions from preliminary results: The pion source 
retains its initial out-of-plane orientation as imposed
by collision geometry, see Fig.~\ref{fig_AZI} (right panel). 
The early pressure and the evolution
time of the system are not sufficient to turn the source
eccentricity into the reaction plane. These results provide
important constraints for dynamical models and may be 
interpreted as an independent indication for a short
evolution time~\cite{lisaismd}. In particular, these findings seem to
disfavor a long-lived hadronic rescattering phase~\cite{teaney}.  

\subsection{Non-identical particle correlations}
Very impressive results on non-identical particle
correlations in Au-Au at 130 and 200 GeV
have been presented by STAR~\cite{kisiel}. 
Such correlations occur due to final state Coulomb and strong
interaction and are sensitive to possible differences of
the freeze-out hypersurfaces of different particle 
species~\cite{lednicky}. 
In the present analysis, correlation functions
for a number of particle combinations have been
presented for the first time ($\bar{p}-\Lambda$,$\pi-\Xi$).
These data can be used to study unknown interaction potentials.

The typical asymmetric correlation patterns have indeed
been observed. 
These asymmetries may point to asynchronous emission 
of different particle species due to sequential freeze-out
or resonance decays. In addition,
in systems with collective expansion,
such differences arise naturally due to space-momentum and
space-time correlations. 
Comparison to a blast-wave approximation 
indicates that strong collectivity leads to asymmetries 
which are consistent with the experimental results.
Further studies need to show whether additional contributions
are indicated by the data.

%{\bf Acknowledgements.} 
%I wish to thank all the speakers 
%for their material 
%and many useful discussions.

\end{document}